\documentclass[preprint2]{aastex}

\newcommand{\gp}{\gamma^\prime}

\newcommand{\tp}{t^\prime}

\newcommand{\Bm}{B_{30}}

\slugcomment{Submitted to the Astrophysical Journal }

\shorttitle{Synchrotron vs.\ EC Models for X-ray Jets}
\shortauthors{Atoyan \& Dermer}

\received{}
\begin{document}

\title{Synchrotron vs Compton Interpretations for 
Extended X-ray Jets}

\author{Armen Atoyan}
\affil{Centre de Recherches Math\'ematiques, Universit\'e de Montr\'eal 
\\Montr\'eal, Canada H3C 3J7}

\email{atoyan@crm.umontreal.ca}

\author{Charles D. Dermer}
\affil{E. O. Hulburt Center for Space Research, Code 7653,\\
Naval Research Laboratory, Washington, DC 20375-5352}

\email{dermer@gamma.nrl.navy.mil}

\begin{abstract}
A widely discussed explanation for the origin of the X-ray emission
observed from knots in extended quasar jets with the {\it Chandra
X-ray Obseratory} is Compton-scattered CMBR by electrons with Lorentz
factors $\gamma^\prime \sim 10^2$.  This model faces difficulties in
terms of total energy requirements, and in explaining the spatial
profiles of the radio, optical, and X-ray knots in sources such as PKS
0637-752, 3C 273, or PKS 1127-145.  These difficulties can be resolved
in the framework of one- and two-component synchrotron models.  We
propose a model where the broad band radio to X-ray synchrotron
emission in quasar jets is powered by collimated beams of ultra-high
energy neutrons and gamma-rays formed in the sub-parsec scale
jets. The decay of the neutral beam in the intergalactic medium
drives relativistic shocks to accelerate nonthermal electrons out of the 
ambient medium.  A second synchrotron component arises from the
injection of leptons with Lorentz factors $\gg 10^7$ that appear in
the extended jet in the process of decay of ultra-high energy gamma
rays. This approach could account for qualitative differences in the
extended X-ray jets of FR1 and FR2 galaxies. Detection of high-energy
neutrinos from blazars and core-dominated quasars will provide
strong evidence for this model.
\end{abstract}

\keywords{galaxies: active --- galaxies: jets --- 
gamma-rays: theory --- radiation processes: nonthermal ---
X-rays: galaxies}  

\section{Introduction }

Three main radiation processes considered to account for the
nonthermal X-ray emission in 
knots and hot spots
 of 
the extended jets discovered by the {\it
Chandra X-ray Observatory} are synchrotron, synchrotron self-Compton
(SSC), and Compton scattering of external photons contributed mostly
by the CMBR (e.g., \citet{hk02,sta03}).  In X-ray knots 
of quasar jets

 with projected lengths $\gtrsim 100$ kpc, where the X-ray
spectrum is not a smooth extension of the radio/optical spectrum, a
currently favored interpretation is the external Compton (EC) model
\citep{tav00,cgc01}. In this model, the X-ray emission from knots such
as WK7.8 of PKS 0637-752 is argued to be due to CMB photons that are
Compton-upscattered by nonthermal electrons from kpc-scale emitting
regions in bulk relativistic motion at distances up to several hundred kpc
away from the central engine.

We note, however, that in the framework of this model it is
problematic to explain a clear trend observed from many extended jets
of radio quasars, such as of 3C 273 \citep{mar01,sam01} or of PKS
1127-145 \citep{siem02}, that the X-ray brightness of the knots
decreases with distance along the jet while the radio flux is {\it
increasing}.  The Lorentz factors of the X-ray emitting electrons in
the EC model are smaller than those of the radio- and optical-emitting
electrons. It is therefore difficult to explain comparable knot sizes
at optical and X-ray frequencies, but extended emission at radio
frequencies.
Moreover, the radiative cooling of these low-energy electrons
is slow, resulting in high total energy requirements.

In this paper, we examine these difficulties and propose an
alternative interpretation for X-ray emission from knots in FR2 radio
galaxy jets and quasars where the radio through X-ray fluxes are
explained by synchrotron radiation from one or two components of
relativistic electrons. Both of these components are powered by
neutral beams of ultra-high energy (UHE) neutrons and gamma-rays
produced by the jet at the base of the central AGN engine
\citep{ad01,ad03}.  The momentum of the decaying beam of neutrons
drives a relativistic shock that accelerates electrons from the
surrounding medium to produce the main nonthermal electron component
responsible for the broad band radio/optical/X-ray synchrotron
emission. In such sources, a second ultra-relativistic lepton
component can also be injected from neutron $\beta$ decay and
the pair production by UHE $\gamma$ rays attenuated by
the CMBR. The synchrotron emission of
these pairs can contribute to or even dominate the synchrotron
X-ray flux in the knots of core-dominated quasars.

Here we focus on knots in quasar jets, though the same 
 2-component synchrotron 
model can apply to hot spots of FR2 galaxies such as Cygnus A
\citep{wys00} or Pictor A \citep{wys01}, where the SSC process could
also play a role. For the X-ray jets in FR1 sources, with projected
lengths of only $\lesssim 5$ kpc, X-ray and optical emission are
generally consistent with a smooth power-law continuation of the radio
spectrum, indicating a synchrotron origin. Mild spectral hardenings at
X-ray energies, such as those observed in the knots of M87's jet
\citep{wy02,mar02}, could still be explained within the context of a
single component synchrotron model (Dermer and Atoyan, 2002;
henceforth DA02).  In FR1 galaxies and BL Lac objects, the neutral
beam power is considerably smaller than in FR2 galaxies and quasars
because of the weaker external radiation field in the inner jet
\citep{ad01}, so that extended jets in FR1 galaxies would primarily
result from the jet plasma expelled directly from the central nucleus.

In Section 2 we discuss difficulties with the EC model, and we propose
one- and two-component synchrotron models for the X-ray jets in
Section 3. The rationale for the model is discussed in Section 4,
including testable predictions from X-ray and high-energy neutrino
observations and a brief discussion of this model in the context of a
scenario for AGN evolution.

\section{Difficulties with the EC Model}

The principal reason for difficulties of the EC model is that the
Lorentz factors $\gamma = E / m_e c^2$ of electrons that produce
X-rays by Compton scattering on CMBR are rather small.  Electrons that
produce synchrotron radio emission at observer-frame frequencies $\nu$
have comoving-frame Lorentz factors $\gp_{syn} \approx 10^3 [(\nu/{\rm
~GHz})(1+z)/\delta_{10}\Bm ]^{1/2}$, where $\delta = [\Gamma (1-\beta
\cos\theta)]^{-1} \equiv 10 \delta_{10}$ is the Doppler factor of the
emitting knot moving with Lorentz factor $\Gamma=(1-\beta^2)^{-1/2}$
at an angle $\theta$ to the observer, and $\Bm \equiv B^\prime/(30\mu$G)
$\gtrsim 1$ is a characteristic magnetic field in the knot derived from
equipartition arguments.\footnote{Primes refer to comoving quantities,
though $B$ is always referred to the proper frame} Electrons that
Compton scatter the CMBR to observed X-ray energies $E_{keV} =
E_{X}/1\,\rm keV$ have $\gp_{\rm C} \approx 90
\sqrt{E_{keV}}/\delta_{10}$.  Consequently the X-ray emitting
electrons will cool more slowly than the radio-emitting
electrons. Converting the comoving Compton cooling time
$t^{\prime}_{C}= t_C/\Gamma\simeq 1.8 \times 10^{12} [(1+z)^4
\gamma^{\prime} \,\Gamma^2 ]^{-1} \, \rm yr$ into the stationary
frame, we find that for the X-ray emitting electrons,
\begin{equation} 
 t_{C} \cong 2.2 \times 10^9 
\left( \frac{\delta}{\Gamma}\right) E_{keV}^{-1/2} (1+z)^{-4} \; 
\rm yr \; . 
\label{tC}
\end{equation}
The coefficient in this timescale corresponds to a propagation
distance of $\approx 630$ Mpc, which is orders of magnitude larger
than the jet extent. This implies a very inefficient extraction of the
injected electron energy.  Note also that the efficiency for Compton
production of the observed X-ray flux cannot be essentially improved
by assuming $\Gamma >> 1$, as suggested earlier \citep{tav00}, since
$\delta \sim \Gamma$ for the energetically most favorable (``not
debeamed'') orientation of the jet, as shown in the next Section.

This leads to large total energy requirements, as well as to a
number of difficulties in the interpretation of spectra and spatial
profiles and sizes of the resolved knots in the radio and X-ray
domains.

\subsection{Total Energy Requirements}

The total electron energy required to produce the $\nu F_\nu$ X-ray
flux $f_X = 10^{-13} f_{-13} \,\rm erg\; cm^{-2}\,s^{-1}$ can be
calculated by noting that the comoving luminosity of a single electron
with Lorentz factor $\gp_{\rm C}$ is $(-dE^\prime/ d\tp)_{\rm C} =
4c\sigma_{\rm T} u_o^\prime \gamma_{\rm C}^{\prime~2}/3$, where
$u_o^\prime = (4\Gamma^2/3) \hat u_{CMB}(1+z)^4$ is the comoving
energy density of the CMBR when $\Gamma \gg 1$, and $\hat u_{CMB} =
4\times 10^{-13}$ ergs cm$^{-3}$ is the local CMBR energy density.
The comoving power $L^\prime_X = N_e\times (-dE^\prime /d\tp)_{\rm C}
\cong 32\pi d^2 \Gamma^2 f_X /\delta^6$ (see eq.[39] in \citet{ds02}),
where $N_e = N_{e}^\prime$ is the total number of relativistic electrons 
with $\gp \sim \gp_{\rm C}$ and $d = 10^{28}d_{28}\,$cm is the luminosity
distance.  This gives for the total energy in the stationary frame
$$W_e \simeq N_e \gp_{\rm C} m_ec^2 \Gamma \simeq $$
\begin{equation}
 2.2\times 10^{61} \; {f_{-13}d_{28}^2\Gamma\over
 \sqrt{E_{keV}}  (1+z)^{4} 
\delta^{5}}\; {\rm erg} \, .
\label{We}
\end{equation}
Note that this derivation  
takes into account the directionality of target photons 
when an isotropic radiation field in the stationary frame
is transformed to the comoving frame \citep{d95,dss97}.
Neglecting this effect gives results accurate 
within a factor of
$\sim 2$ when $\delta \approx \Gamma$, but can introduce 
large errors when $\delta \ll \Gamma$.

The energy given in eq.\ (\ref{We}) accounts only for electrons
with $\gp \sim \gp_{\rm C}$, and neglects protons normally required to
ensure neutrality.  Since the proton energy even at rest is $\gtrsim
10$ times larger than the comoving energy of electrons with $\gp_{\rm
C} \sim 100$, the total energy of particles $W_{tot} \gtrsim 10 \times
W_{e}$ is needed, unless $ e^{+}-e^{-}$ pair-dominated plasma could be
formed in the knot. \citet{cf93} argue, however, that quasar jets are
composed of e-p plasma by comparing jet luminosities inferred from SSC
models with the radio lobe powers.

The total energy requirements can be reduced to values reasonable for
X-ray knots with a typical size $R_{kpc} = R/\,\rm kpc \sim 1$
only by assuming $\delta \sim \Gamma \gtrsim 10$.  For a jet moving at
an angle $\theta$ with respect to the line-of-sight, the maximum
Doppler factor is $\delta_{\rm max} = 1/\sin \theta$, which is reached
if also $\Gamma = 1/\sin\theta$.  This implies very small inclination
angles of jets up to $\theta \leq 1/\Gamma \lesssim 5^\circ$ in order
to have $\delta \gtrsim 10$.

For knot WK7.8 in PKS 0637-752 at $z = 0.651$, $d_{28} = 1.20$
(for a $\Lambda$CDM cosmology with $\Omega_m = 0.3$, $\Omega_\Lambda =
0.7$ and Hubble constant of 70 km s$^{-1}$ Mpc$^{-1}$) and $f_{-13}
\approx 0.4$ (from \citet{sch00}, including bolometric correction for 
the integral flux), equation (\ref{We}) results in 
$W_e\simeq 2\times 10^{56}\; \rm erg$ and $W_{tot} \gg 10^{57}\,\rm erg$ 
in the stationary frame, if one assumes
$\theta \simeq 5^\circ$ and $\Gamma \approx \delta \cong 10$.

For 3C 273 at $z=0.158$ and $d_{28} = 0.23$, the angle $\theta \cong
17^{\circ}$ explains superluminal motion of the sub-pc jet
\citep{rtr94}, and it may be as large as $\simeq 30^\circ$--$35^\circ$
for the kpc-scale jet as deduced from polarization measurements by
\citet{cond94}. Even taking $\theta = 17^\circ$ for knot A1 with
measured bolometric flux  $f_{-13} \simeq 3$ in the keV region
\citep{mar01}, the minimum electron energy when $\Gamma = 
\delta = 3.4$ is $W_{e} \gtrsim 1.4 \times 10^{58} \; \rm erg$ for the Chandra
observations. This implies a total energy $W_{tot} \gtrsim 10^{59}$
ergs in the stationary frame, or $W_{tot}^\prime \simeq W_{tot}/\Gamma
\gg 10^{58}$ ergs in the comoving frame.

For an X-ray knot size $R_{kpc}\sim 1$, an equipartition magnetic field 
for the comoving energy $W_{tot}^{\prime} = 
10^{56} \,W^{\prime}_{56}\,  \rm ergs$
is $B_{eq}^\prime \cong 150 \, (W^{\prime}_{56})^{1/2}
R_{kpc}^{-3/2} \; \rm \mu G$. 
For knot WK7.8 in PKS 0637--752, the size $R_{kpc}\approx 1 \,\rm kpc$
 corresponding to FHWM $\Delta \theta = 0.3^{\prime\prime}$ (but not 
$R\simeq 3\,\rm kpc$, as in \citet{tav00}, if one takes into account the
 redshift effect, $R = d \, \theta /(1+z)^2 $) and $W_{56}^\prime \approx 0.2$
in case of $\delta = \Gamma = 10$.
This results in the equipartition magnetic field 
$B_{eq}^\prime \approx 70 \,\rm \mu G$ (instead of $\simeq 15 \,\rm \mu G$
for 3$\times$ larger knot size).
This field is unacceptably high for the EC model because it would
overproduce the observed radio flux, given the number of electrons
with $\gamma^\prime\sim \gamma^\prime_{C}$ needed for the X-ray
flux. However, in case of magnetic fields at the level of only a few
tens of $\rm \mu G$, the energy density, and thus the pressure, of
relativistic particles would be much higher than the pressure of the
magnetic field.  Therefore either the confinement time of electrons in
the knot would be $t_{conf}^\prime
\sim R/(c/\sqrt{3}) \simeq 2\times 10^{11} R_{kpc}\, \rm s$, or else
the knot would be inflating on the same timescale. This implies a
comoving injection power in the knot $L^\prime \gtrsim 5\times 10^{44}
W^{\prime}_{56}/R_{kpc} \,\rm erg \, s^{-1}$, or a stationary-frame
injection power $\Gamma^2 L^\prime \gtrsim 10^{47}$ ergs s$^{-1}$. In
fact, taking into account contribution of protons,
\citet{tav00} require a power of $3\times 10^{48}$ ergs
s$^{-1}$ to model the spectral energy distribution (SED) of knot WK7.8
of PKS 0637-752 with the EC model. Although they argue that this is
consistent with total kinetic power of jets inferred to power the
giant radio lobes of radio galaxies, the largest jet power in the
sample of \citet{rs91} hardly exceeds $10^{47}$ ergs s$^{-1}$. This
power also is 1--2 orders of magnitude greater than the maximum peak
$\gamma$-ray luminosities inferred from EGRET observations of blazars,
taking into account the likely beaming factor of $\sim 1$\% \citep{mc99}.

It should be noted here that in principal it is possible to satisfy the
equipartition condition also for a given size $R=1 \,\rm kpc$ of the knot 
WK7.8, increasing the Doppler factor $\delta$ further to $ \simeq 20$ 
\citep{da04}. This will also minimize ({\it somewhat reducing}) the power 
requirements for the EC model \citep{gc01,da04}. However, this is
possible only if the jet inclination angle is further decreased to
$\leq 3^\circ$.  Although this cannot be excluded for a particular
source, such angles would be problematic to assume for many sources,
as we discuss in Section 4.

\subsection{ X-rays vs.\ Radio Spatial and Spectral Profiles }

In the EC/CMBR model, X-rays are produced by electrons with
$\gamma_{\rm C}\sim 10^3 $ in the stationary frame.  They cool on
the CMBR only on Gyr timescales $t_{cool} \simeq t_{C}$ of eq.\,(\ref{tC}),
 unless their overall cooling is dominated by synchrotron
losses, which would then overproduce the observed radio flux at low
frequencies. Since the photon target for the Compton process does not
degrade at any distance $r$ from the parent AGN, the strong fading of
the X-ray knots and jet generally (outside the knots) with $r$ in many
FR2 quasars, such as in 3C 273 \citep{sam01,mar01}, PKS 1127-145
\citep{siem02}, or Pictor A \citep{wys01}, implies either a
 fast decrease of the total number of electrons with $\gamma \sim
\gamma_{\rm C}$ or a decline of the jet Doppler factor. Because the
observed radio emission is produced by electrons with $\gamma_{syn}$
larger than $\gamma_{\rm C}$ only by a factor $\lesssim 10$, both
assumptions would imply a rapid fading of the radio brightness with
distance as well. Meanwhile, exactly the opposite radio brightness
profiles are observed.

Secondly, because the cooling time is shorter for radio than for
Compton X-ray emitting electrons, one should also expect that the
X-ray knots in the EC model would be more diffuse than the radio or
optical knots. Again, this prediction is in disagreement with
observations. Note in this regard that the adiabatic cooling of X-ray
emitting electrons due to fast expansion of X-ray knots, as proposed
recently \citep{tgc03} for the interpretation of fading of the X-ray
brightness outside the knots, would also equally impact the 
radio-emitting electrons \citep{sta04}. Therefore this scenario would not
resolve either of these two discrepancies between the X-ray and the
radio brightness patterns.

One more difficulty for the Compton interpretation arises for those
jets where the spectral indices $\alpha$ (for the spectral flux
$F_{\nu} \propto \nu^{-\alpha} $) are significantly different at radio
and X-ray frequencies. As also discussed earlier, e.\ g., by
\citet{hk02}, spectral profiles steeper in X-rays than in radio,with
$\alpha_X > \alpha_r$, and especially jets/knots 
in a number of FR1 galaxies 
where $\alpha_{X} \gtrsim 1$, represent strong evidence for the
synchrotron origin of the X-ray flux, implying acceleration of 
electrons with $\gamma^\prime \gg
10^7\sqrt{E_{keV}(1+z)/\delta_{10}B_{30}}$ in the knots of radio
jets. This also is the case for many terminal hot spots, such as the
western hot spot of Pictor A with $\alpha_{r} \approx 0.74 $ and
$\alpha_X = 1.07 \pm 0.11$ \citep{wys01}.

It should be pointed out here that these difficulties with the
interpretation of different spectral and opposite spatial profiles in
radio and X-ray patterns of jets relate not only to EC models, but
equally to SSC models.  The X-ray flux in SSC models is likewise
produced by low-energy electrons that emit the synchrotron radio
photons for subsequent Compton interactions, and the density of this
target does not decline, and may even increase, along the jet.

\section{Synchrotron Origin of X-rays}

In the X-ray jets of FR1 galaxies, such as 3C 66B \citep{hbw01} and
some knots in M87's jet \citep{wy02,mar02}, the X-ray--optical--radio
spectra are consistent with a single or smoothly steepening power-law
spectrum, which is readily explained by a single-component synchrotron
spectrum. It is a common pattern that the peaks of the emission of the
knots and hot spots in the radio, optical, and X-ray domains of quasar
jets are well correlated and often coincide, such as in the jet of 3C
273 \citep{sam01,mar01}. When the profiles do not coincide, as in the
FR1 galaxy M87 \citep{wy02} or the GPS quasar PKS 1127-145 \citep{siem02},
the peak emission of the X-ray profile is closer to the core than the
radio profile.

This behavior suggests a common synchrotron origin for the radiation
in all bands, where nonthermal electrons accelerated at a shock lose
energy through adiabatic and radiative processes as they flow
downstream from the shock. The synchrotron optical and X-ray emissions
from the highest energy electrons have roughly coincident profiles,
but the radio emission is far more extended due to the longer cooling
timescale, and can even be offset as a consequence of a low-energy
cutoff in the electron injection function.\footnote{\citet{gk04}
explain this offset in the context of an EC model where the flow
decelerates downstream from the injection site, but must require that
the magnetic field is amplified through compression in the
downstream decelerating flow.}

In apparent conflict with a synchrotron model are the SEDs of many
quasar knots and hot spots that show a general behavior whereby the
optical spectra are steeper than both the radio and the X-ray spectra,
and the X-ray fluxes are above the extrapolation from the optical
band. This behavior cannot be immediately explained by a ``standard''
synchrotron model with a single power-law electron component. It
apparently requires a second electron component at very high energies
$\gtrsim 10$ TeV with a sharp cutoff at lower energies.
However, the origin of such electrons in the knot was not easy to
understand and obviously needs an explanation.

\subsection{Single-component model}

In our model (DA02), we have shown that in many cases even a single
population of electrons with a power-law injection spectrum
$Q_{inj}(\gamma^\prime) \propto \gamma^{\prime -p}
e^{-\gamma\prime/\gamma^\prime_{\rm max}}$ could explain the observed
spectral peculiarities if the electrons are accelerated to
sufficiently high energies with $\gamma^\prime_{\rm max} =
E^\prime_{\rm max} /m_e c^2 \gtrsim 10^8$.  When CMBR cooling in the
Thomson regime exceeds synchrotron cooling, i.\ e., when the magnetic
and radiation field energy densities relate as $u_o^{\prime} =
\hat u_{CMB} (1+z)^4 \,(4\Gamma^2/3) \gtrsim u^{\prime}_{B}=(B^\prime)^2
/8\pi$, a hardening in the electron spectrum is formed at comoving
electron energies $\gamma^\prime \gtrsim 10^8/\Gamma (1+z)$, where KN
effects in Compton losses become important. The spectrum steepens
again at higher energies when synchrotron losses take over.  For
$B^\prime$ up to few tens of $\mu \rm G$, this condition is satisfied
for jets with $\Gamma \gtrsim 10$.  Such an effect produces a hardening
in the synchrotron spectrum between optical and X-ray frequencies.

\begin{figure}[t]
\plotone{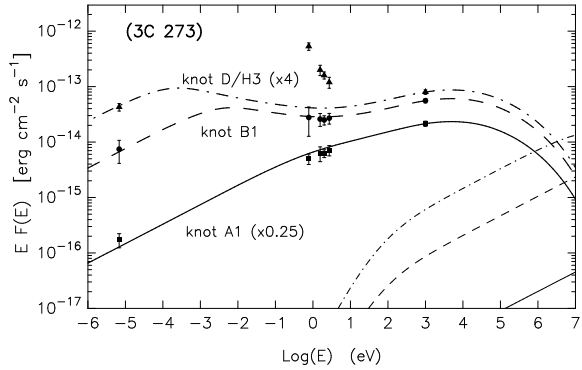}
\caption{ SED fits to the broad band fluxes of knots A1, B1,
and D/H3 in the jet of 3C 273, calculated in the framework 
of single-component synchrotron
model of \citet{da02}. For knots D/H3 and A1, the measured fluxes 
\citep{mar01} are displaced up and down by factor 4, respectively,
 to avoid overlaps in the optical band. The electron injection rates
provide the observed X-ray fluxes at 1 keV.  
 For all knots most of model parameters are assumed the same (see text): 
electron injection spectrum with
$p = 2.3$ and high and low cutoff energies $\gamma_{\rm max}^\prime =
2\times 10^8$ and $\gamma_{\rm min}^\prime = 2\times \Gamma$, 
$B =15\,\rm \mu G$, $\Gamma_{\rm knot} = 15$,
$\theta_{\rm jet} = 17^\circ$ (resulting in $\delta_{\rm knot} \simeq
1.5$). The main difference is in different escape (or else effective
injection) times for the various knots, namely $t^{\prime}_{\rm
esc.A1} = 1.6\times 10^3 \,\rm yr$, $t^{\prime}_{\rm esc.B1} =
3.8\times 10^4 \,\rm yr$, and $t^{\prime}_{\rm esc.D/H3} = 1.5 \times
10^5 \,\rm yr$.  The thin solid, dashed and dashed-dotted curves show
the fluxes of EC radiation on CMBR for these parameters of knots.  }
\end{figure}

In Fig.\ 1 we show fits to the multiwavelength SEDs of knots A1 and B1
of the jet of 3C 273, with deprojected length $> 100 \,\rm kpc$,
calculated in the framework of a single-component synchrotron model
(DA02). For all knots, $B = 15\,\rm \mu G$ (including the knot
D/H3), and practically all other model parameters are also the same.
The single parameter which is different for the 3 curves is the
time from the start of electron injection, which can also be
interpreted as the different escape times of electrons from the knots.
These times are given in the figure caption.  The curves are 
are normalized to the observed X-ray fluxes at 1 keV.  
The total electron
energies in knots A1 and B1 are $W_{e.A1}^\prime =1.3\times
10^{56} \,\rm erg$ and $W_{e.B1}^\prime =1.6\times 10^{57} \,\rm erg$,
and the required electron injection powers are $L_{e.A1}^\prime
=2\times 10^{45} \,\rm erg/s$ and $L_{e.B1}^\prime =1.3\times 10^{45}
\,\rm erg/s$.

Note that the spectra in Fig.~1 are fitted for the jet inclination
angle $\theta = 17^\circ$ in 3C 273. For the assumed Lorentz factor
$\Gamma= 15$, this angle is much larger than the effective beaming
angle of the inner jet emission $\theta_{beam}\cong 57^\circ/\Gamma$,
and results in $\delta \approx 1.5 \ll \Gamma$.  Even for this
strongly debeamed case, the simple ``one-zone and single-component''
synchrotron model provides good fits to the observed fluxes of the
knots A1 and B1 with still acceptable total energy requirements.  The
thin curves in Fig.\ 1 show the fluxes to be expected from the EC
process for the same model parameters. It follows from comparing these
fluxes with the observed X-ray fluxes that for the given parameters of
3C 273, the EC model would require 2-3 orders of magnitude more total
energy for both knots A1 and B1. Even formally assuming an
unreasonably large injection/escape time, however, it is not possible
to explain the SED of the knot D/H3 in the framework of a
single-component synchrotron model.

\subsection{The second electron component}

The X-ray emission in knot D/H3 of 3C 273, or generally in knots
where either the optical flux is very steep, $\alpha_{\rm opt} \gtrsim
1.5$, or the energy flux $\nu F_{\nu}$ is much higher in X-rays than
in the optical band, can only be explained by synchrotron radiation
either as a second component of electrons or, as another possibility,
synchrotron emission of UHE protons with $\gamma_p
\gtrsim 10^9$ and magnetic fields as strong as $B_{\rm mG} =
B/10^{-3}\,\rm G \gtrsim 1$ \citep{aha02}.

The energy requirements of the proton synchrotron model are as
demanding as of the Compton model. In this model, keV radiation is
produced by protons with comoving Lorentz factors $\gamma^{\prime}_{p}
\simeq 4\times 10^{8} \,[E_{keV} (1+z)/B_{\rm mG}^{\prime}
\delta]^{1/2}$.  The synchrotron cooling time of these protons
$t_{s.p}^\prime \simeq 1.5 \times 10^{17}/
( B_{\rm mG}^{2}\gamma^{\prime}_{p}) \,\rm yr$ which, in
the stationary frame, is \begin{equation} t_{s.p.} = 3.8 \times 10^8 \,
\Gamma \delta^{1/2} E_{keV}^{-1/2} B_{\rm mG}^{-3/2} (1+z)^{-1/2} \;
\rm yr\, .
\label{tsp}
\end{equation}
Comparing this time with eq.\ (\ref{tC}) shows that the energy
required in UHE protons alone in the knot should be practically as
large as in electrons with $\gamma^\prime \sim \gamma^\prime_C $ in
the EC model. At the same time, one also has to assume that only a small
amount of energy is injected in radio electrons in order not to
overproduce the observed radio flux in the high magnetic field in the
knot.

The origin of the second electron component to produce X-ray
synchrotron radiation in a leptonic model may, however, seem ad hoc.
Noticeably, this second electron component should be concentrated
mostly at multi-TeV and higher energies,  with a cutoff at energies
below $\gamma^\prime\sim 10^7$, or otherwise the synchrotron flux of
this component would exceed the observed flux in the optical
band. After discussing in this Section the requirements and demonstrating the 
feasibility of a two-component
synchrotron model, we propose a scenario for 
the origin of the second component  in Section 3.3.

In Fig.~2 we show a fit to the SED of knot WK7.8 of PKS 0637-752,
calculated in the framework of a 2-component synchrotron model.  We assume
a power-law spectrum with $p_1 = 2.6$ to fit the radio spectrum, and an
exponential cutoff above $\gamma^\prime_{1.max} =1.8 \times 10^5$, which
explains the steep optical flux.  The injection spectrum of the second
electron component has a power-law index $p_2=2.2$, with exponential
cutoffs below $\gamma_{\rm 2.min}^\prime =2\times 10^7$ and above
$\gamma_{\rm 2.max}^\prime= 10^{11}$.  The synchrotron radiation of
this second electron component is shown by the heavy dashed curve
for an escape time $t^\prime_{\rm esc}=1\,$kyr. For comparison, the
thin short dashed curve shows the evolution of the fluxes in case of a
larger escape time of the electrons from the knot, with $t^\prime_{
\rm esc}=3\,$kyr.  As can be seen,  further increase of this parameter
beyond $t^\prime_{ \rm esc}\simeq 10^4\,$yr when 
$B^{\prime}=100\,\rm \mu G$ would result in the
synchrotron flux of the rapidly cooling second electron component
exceeding the observed flux in the optical band.

\begin{figure}[t]
\plotone{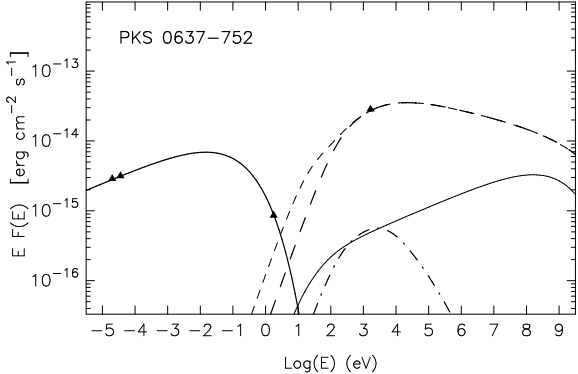}
\caption{ Spectral fits to the fluxes of knot WK7.8 
of PKS 0637-752 \citep{sch00,cha00}, calculated in the
framework of a two-component synchrotron model, assuming $B^\prime =100
\,\rm \mu G$, and $\Gamma = 5$ and $\theta_{\rm jet} =
10^\circ$ for the jet.  The heavy solid curve shows the synchrotron
flux from the first electron component with the injection index
$p_1 =2.6$, the exponential cutoff above $\gamma_{\rm 1.max}^\prime =
1.8\times 10^5$, and a cutoff below $\gamma_{1. \rm
min}^\prime \simeq 20 $. Thin solid curves show Compton fluxes from these
electrons.  For the second electron
component, an injection spectrum with $p_2 = 2.2$ at energies
$ 2\times 10^7 \lesssim \gamma^\prime \lesssim 10^{11}$, and with
cutoffs outside this interval is assumed.  The heavy dashed
line shows synchrotron radiation of this component, and the dot-dashed
curve shows the radiation from the 
pair-photon cascade initiated by the second electron component in the
knot. Electron escape time $t^\prime_{ \rm esc} = 1 \,\rm
kyr$ is assumed. For comparison, the short thin dashed curve shows 
the flux of the second component in case of 
$t^\prime_{ \rm esc} =3 \,\rm kyr$. }
\end{figure}

Longer escape/injection times $t^\prime_{ \rm esc} \gtrsim
3\,\rm kyr$ are allowed for smaller magnetic fields $\lesssim 100\,\rm
\mu G$ in knot WK7.8.  The interpretation of these timescales when
$t^\prime_{ \rm esc} \sim 3$-$10 \,\rm kyr$ suggests that X-ray knots
are manifestations of relativistic shock waves, with transverse size
scales of kpc dimensions, that are moving towards us (though generally
at some angle $\theta \sim 10^\circ $-$20^\circ$) with $\Gamma
\sim 10$. These shocks accelerate electrons and can amplify the
magnetic field in the downstream flow. In the shocked material, the
relativistic plasma expands at speeds close to $ c/\sqrt{3}$ (in the
shock frame).  Then the escape times $t^\prime_{ \rm esc}
\sim 3-10 \,\rm kyr$ would imply an effective thickness of the region with
enhanced magnetic fields downstream of the shock equal to about $h
\simeq (c/\sqrt{3})\times t^\prime_{\rm esc} \sim 1 \,\rm kpc$. This is
comparable with the given transverse size of the shock, and therefore 
even a smaller escape times could be still reasonable.

It is important that the energy requirements derived for the first and
 second electron components are low.  For the assumed parameters of
 knot WK7.8, the total comoving energies in the first
 components is $W_{e.1}^\prime = 3.9\times 10^{55}
\rm \, erg$, and only $W_{e.2}^\prime = 5.5 \times 10^{51}\,\rm erg$ in the
second component.  The energetics $W_{e.1}^\prime$ of the first
component is contributed mostly by radio-emitting electrons which cool
slowly and can be accumulated in a larger region over a longer time
than $t^\prime_{\rm esc} \simeq 1\,$kyr assumed in Fig.~2.  Comparing
the EC flux shown by the thin solid line with the observed X-ray flux,
one can see that the chosen parameters for knot WK7.8 would require in
the EC model a total energy budget larger by 2 orders of magnitude
than found in Fig.~2, namely $W_{e}\simeq \Gamma W_{e}^\prime >
10^{58}\,\rm erg$. It would also require a magnetic field well below
the near-equipartition value assumed in Fig.~2 in order to reduce the
radio emissivity.

For the fast-cooling second component of electrons, a more
informative measure than $W_{e.2}$  is the comoving injection power,
or injection ``luminosity'' $L_{e.2}^\prime= 10^{43} \, L_{43}^\prime
\,\rm erg \, s^{-1}$.  The model fit to the observed X-ray flux in
Fig.~2 requires $L_{43}^\prime = 0.3$. This luminosity implies that
 electrons with average Lorentz factor
$\gamma_{8}^{\prime} = \gamma^{\prime}/10^8 \simeq 1$ are injected
at the rate 
\begin{equation}\label{Ndot}
\dot{N}_{e.2}^\prime \simeq 10^{41} \,L_{43}^\prime 
(\gamma_{8}^\prime)^{-1} \; \rm s^{-1} \; .
\end{equation}
Because of relativistic time contraction $dt^\prime = dt/\Gamma$,
this injection rate becomes even smaller in the stationary frame,
 $\dot{N}_{e.2} = \dot{N}_{e.2}^{\prime}/ \Gamma$.

The steep spectra of the first electron component with power-law
injection index $p=2.6-2.7$ needed to explain the radio fluxes of the
knot WK7.8 imply that the shock is not very strong. This is in a good
agreement with the relatively low value of the characteristic maximum
energy for the first electron component which is $\gamma_{\rm
1.max}^\prime = 1.8\times 10^{5}$ in Fig.~2.  But then the production
of a second electron component reaching much higher energies appears
to be problematic.

\subsection{Model for the Second Electron Component}

The problem is solved if the second electron component is not
accelerated by the shock, but rather is swept up in the upstream
plasma. Spectra of electrons of the type required, with a sharp cutoff
below multi-TeV energies, could appear in knots at distances from a few
tens to $\gg 100 \,\rm kpc$ from the decaying neutral beam of UHE
neutrons and gamma-rays produced in the relativistic compact jet of
AGN at sub-pc scales, as we have recently suggested \citep{ad01,ad03}.
The total power transported by such a beam in FR2 quasars to distances
0.1-1 Mpc can be very significant.  The energy released in UHE
neutrons at energies $\sim 10^{17}$-$10^{20}\,$eV can reach a few to
several per cent of the entire energy of protons accelerated by the
inner jet \citep{ad03}.

The total number of electrons in the second component is so small that
even the number of $\beta$-decay electrons from the decay of neutrons could
be sufficient for eq.\ (\ref{Ndot}). The decay lifetime of neutrons at
rest is $\tau_0 \simeq 900\,\rm s$, so that neutrons with Lorentz
factors $\gamma_n\gtrsim 10^8$ decay on kpc distances from the centre.
The spectrum of neutrons at distance $r = r_{\rm kpc} \,\rm kpc$ is
\begin{equation}\label{Nr}
N_{r}(\gamma) = N_{0}(\gamma) \exp(- \frac{r}{c\tau_0 \gamma_n}) =
 N_{0}(\gamma) \exp (- \frac{1.1 r_{\rm kpc}}{\gamma_{n.8}}),
\end{equation}
where $N_{0}(\gamma)$ is the spectrum of neutrons produced in the
compact jet at $\lesssim$ parsec scales. For a power-law injection 
spectrum of neutrons given by
$N_{0}(\gamma_n)
\propto \gamma_{n}^{-2},$ the intensity of neutron-decay electrons at distance
$r$ is equal to
\begin{equation}\label{Nbeta}
\dot{N}_{\beta}(r)\simeq 2.7\times 10^{46} r_{\rm kpc}^{-2} \, W_{n.56}
\; \rm s^{-1}\; ,
\end{equation} 
where $W_{n.56}=W_{n}/10^{56}\,\rm erg$ is the total energy in the
neutrons.  The number of beta decay electrons $\dot{N}_{\beta}$ could
suffice in eq.\ (\ref{Ndot}) up to $r \simeq 1\,\rm Mpc$ distances,
provided that the spectrum of neutrons in the neutral beam extends to
$10^{20}\,\rm eV$.  Note that the Lorentz factors of $\beta$-decay
electrons are about the same as the Lorentz factors of parent
neutrons, i.\ e., $\gamma_e \gtrsim 10^{8}\,\rm r_{kpc}$.

If an amount of energy $W_{n}$ in UHE neutrons is ejected from the
central engine during a period $\Delta t_{high}\simeq  3\,\rm kyr$
of high-state activity, then these neutrons will form a slab of
$\simeq 1 \rm \, kpc$ length moving along the jet at nearly the speed of
light.  The production power of UHE neutrons $L_{n} \simeq
W_{n}/\Delta t_{high} \simeq 10^{45} \,W_{n.56}\,\rm erg/s$ could be
provided if the total power of acceleration of protons in the AGN core
would be $\sim 3\times 10^{46} \,W_{n.56}\,\rm erg/s$. This is quite 
acceptable for the powerful central engines of FR2 galaxies \citep{rs91}.

The power requirements on the central engine are significantly
more relaxed if we take into account $e^+$-$e^-$ pairs produced in
photoabsorption of the $\gamma$-ray component of the UHE neutral beam.
These electrons are produced with Lorentz factors $\gtrsim 10^{10}$ at
distances $\gg 10\,\rm kpc$ (see Atoyan \& Dermer, 2003).  Because of the
relativistic Klein-Nishina (KN) effect, the efficiency of the
electromagnetic cascade of these leptons on the CMBR is essentially
suppressed in an ambient magnetic field as low as 
$\gtrsim 1\,\rm \mu G$. Indeed, the Compton cooling time 
of electrons with $\gamma_{10}\equiv \gamma/10^{10} \gtrsim 1$ on CMBR is 
$t_{\rm KN} \simeq 5 \times 10^4 (1+z)^{-2} \gamma_{10} \,\rm yr$, while the
synchrotron cooling occurs on timescales $t_{syn} \simeq 1.6 \times 10^3 
(B_{\perp}/ {1 \,\rm  \mu G })^{-2} \gamma_{10}^{-1} \,\rm yr$. In order
to have an efficient cascade, \citet{ner02} have assumed a jet with
magnetic field of order $\geq 10 \,\rm \mu G$ (which is needed for
effective synchrotron radiation), but which is extremely well aligned,
such that the perpendicular component of the field is very small,
$B_{\perp}\lesssim 0.1\,\rm \mu G$.

In our model, where the flux detected from X-ray knots is produced in
the downstream regions of the forward shocks, we can allow
significantly less ordered magnetic fields in the upstream region of
the shock, with $B_{\perp} \sim 1 \,\rm \mu G$. In this case the
efficiency of Compton emission which defines the efficiency of the
cascade of pairs with initial
Lorenz factors $\gamma \gtrsim 10^{10}$ would be only 
$ t_{syn}/t_{\rm KN} \lesssim 10^{-2}$.  
The energy of injected UHE pairs, however, is orders of
magnitude larger than the energy $\gamma\gtrsim 10^{8}$ of $e^+$-$e^-$
pairs effectively produced {\it in} the cascade. Furthermore, the efficiency
of the Compton process dramatically rises when the synchrotron losses
bring the injected electrons (of both signs) closer to $\gamma \simeq 2\times
10^9$. This implies that the number of electrons produced from a
single initial UHE $\gamma$-ray photon could be significantly larger
than the number of $\beta$-decay electrons.  Therefore in case of
comparable powers in the neutron and gamma-ray components of the
neutral beam, as expected for FR2 radio quasars \citep{ad03}, the total
{\it number} of leptons with $\gamma \gg 10^7$ needed for the second
electron population in the knots would be determined by the $\gamma$-ray
component of the UHE neutral beam.
 
In the stationary-frame magnetic field $B_{up.\perp} \gtrsim 1 \, \rm
\mu$G upstream of the shock (corresponding to comoving
$B^{\prime}_{up.\perp} \sim
\Gamma \times B_{up.\perp} \gtrsim 10\,\rm \mu G$) 
these electrons cool to energies $$\gamma_{min} \simeq 4 \times
10^{10} (\Delta t_{kyr})^{-1} (B_{up.\perp}/1\,\rm \mu G)^{-2} $$
during time $\Delta t = \Delta t_{kyr} \,\rm kyr$ since their
production in the jet fluid and until they are overtaken by the
trailing relativistic shock. Note that our model does not require an
extremely ordered magnetic field along the jet, although stretching of
 the pre-existing intergalactic magnetic field along the jet
  is a very plausible outcome of the decaying neutral beam and appearance 
of the beam of charged UHE secondaries driving the extended jet.  
Because the time of injection
for different electrons would be different, a broad spectrum of
ultrarelativistic $e^+$-$e^{-}$ pairs up to their energy at
production, will be overtaken.  The enhanced magnetic field downstream
of the shock results in rapid radiative losses, a ``lighting up''
these electrons and the appearance of an X-ray knot.

In case of $\gamma_{2.\rm min} \gg 10^8$, the cooling synchrotron
spectrum with $\alpha_{X} \simeq 0.5$ is produced. A small
contribution of electrons from the pair-photon cascade on the CMBR, as
discussed above, could effectively result in $\gamma_{\rm 2.min}
\lesssim 10^8$.  It may also be the result of higher $B_{up.\perp}$
and longer cooling time $\Delta t$, which in principle could reach
$\gtrsim 10^5\,$yrs at $>100\,\rm kpc$ distances.  In that case
steeper synchrotron spectra at keV energies will be produced, as in
Fig.~2.

\section{Discussion}

The external Compton model for extended X-ray jets faces difficulties
with large energy requirements, and in explaining the different
spatial profiles at radio, optical, and X-ray frequencies. It also
cannot be invoked to explain knots and hot spots where the X-ray spectral
indices are significantly steeper than the radio spectral indices, 
in which case a synchrotron model is favored.

The large energy requirements in the EC model can be reduced to
acceptable values only by assuming Doppler factors $\delta \geq
10$. In these cases, the jet would have to be directed towards us at
very small angles of only $\theta \leq 6^\circ $.  Although we cannot
exclude that the number of X-ray jets with such small angles could
still be significant, the probability for one of the 2 jets of any
quasar to be directed within such an angle is only $P_\theta = (1-\cos
\theta)/2\pi \leq 8.7\times 10^{-4} (\theta/6^\circ)^2$. Allowing for
X-ray jets with $\theta \lesssim 20^\circ$ will increase the
probability to detect several such sources by several orders of
magnitude.  A common model for X-ray knots would then have to 
be able to deal with
sufficiently large $\theta$ and to allow debeamed jets with
$\delta \ll \Gamma \simeq 10$.  The energy demands for such jets
become unrealistic in the EC model.

The large energy requirements are significantly relaxed
for synchrotron models. Synchrotron X-rays are produced by very high 
energy electrons with short cooling times, which minimizes
the injection power of electrons needed to explain the X-ray flux. 
 For a two-component synchrotron model, unlike for the EC model,
the magnetic field is not limited by the ratio of X-ray to radio
fluxes and a given inclination angle (obtained, for example, by fixing
the maximum possible Dopler factor). This
minimizes the total energy
accumulated in the form of radio emitting electrons and magnetic fields
in the knot needed to explain the observed radio flux.  
The comoving equipartition
fields in FR2 knots are typically at the level $\sim 50$--$100\,\rm \mu G$,
which are higher than $B \sim 10$-$20 \,\rm \mu G$ deduced 
from EC models. The synchrotron interpretation of X-rays also allows 
one to have significantly debeamed jets
and still remain within an acceptable range for the energy budget,
as in the knots of 3C 273 (Fig.~1).

The relative advantage of synchrotron models for debeamed jets with
$\delta << \Gamma$ is due essentially to the more narrow beaming
diagram for the EC radiation than for the synchrotron radiation in the
stationary frame \citep{d95}. This is the case when the photons are
detected at an angle $\theta \gg 1/\Gamma$. In
the comoving frame such photons are produced in Compton `tail-on'
collisions with beamed (in the comoving frame) external photons at
angles $(1-\cos \theta^\prime)\ll 1$; therefore the efficiency of the
Compton process is strongly reduced. In the meantime, synchrotron
radiation production in a quasi-isotropic knot magnetic field is
isotropic in the comoving frame.

The steeper SED in the optical than X-ray bands in some knots, as A1
or B1 knots of 3C 273 in Fig.~1, can be explained in the framework of a
single population of electrons accelerated to multi-TeV energies in
the shocks forming the knots (DA02). The two-component synchrotron
model appears more general. In particular, it can explain practically
all types of SEDs detected from the knots and hot spots of multi-kpc
scale jets of FR2 radio galaxies, such as the knot WK7.8 of PKS
637-752 where the optical flux is a factor $>10$ times below the
power-law extrapolation of fluxes from the radio to X-ray bands.

We have argued that the origins of the two different components in the
knots and hot spots of FR2 jets can be naturally explained within a
unified model for large scale jets as manifestations of neutral beams
\citep{ad01,ad03}.  These beams, composed of UHE neutrons, 
gamma-rays and neutrinos, are efficiently produced in the compact
relativistic jets of FR2 quasars by accelerated UHE
protons in the process of photomeson interactions with the accretion disk
radiation on $\lesssim$ pc scales. 
The decay of UHE neutrons deposits momentum and energy of the
neutral beam into the intergalactic medium, and drives surrounding
plasma to relativistic motion (presumably with moderately relativistic
speeds, though this will require a hydrodynamic study to quantify)
through interactions with ambient magnetic fields and the generation of
plasma waves. 

The appearance of knots in continuous jets could be manifestations of
shocks resulting from the increased activity of the central engine
during some time in the past \citep{ner02,ad03}. The first population
of electrons is produced in the process of first-order Fermi
acceleration of electrons from the surrounding intergalactic
medium. The radio electrons cool slowly; therefore they drift and
accumulate along the jet, which explains the increase of radio
brightness.

A second population of electrons in the jet originates in the  
decay of UHE gamma-rays with Lorentz factors
$\gamma_0 \gtrsim 10^{10}-10^{13}\,\rm eV$
\citep{ad03}, and can be contributed even by $\beta$ electrons from neutron
decay. It is essential that the production spectrum of these electrons
is cut off at lower energies, $\gamma_c \lesssim \gamma_0 $. Because
of the strong Klein-Nishina effect for the Compton radiation of UHE
electrons, most of the electron energy will be deposited in the form
of synchrotron radiation at MeV/GeV energies in ambient fields $B
\gtrsim 1 \mu$G. This radiation can be effectively lost for an
observer outside of the jet opening angle. These electrons will be
cooled down to energies $\gamma \gtrsim 10^{8}-10^{9}$ before being
overtaken by the relativistic shock.  Reprocessing of even a small
fraction of the $\gamma$-ray beam energy in the pair-photon cascade
along the jet, however, will result in a very significant contribution
to the total number of UHE electrons, with a broad spectrum above
$\gamma \sim 10^8$ and a cutoff at lower energies. These electrons are
deposited throughout the length of the jet, and will be found, in
particular, in the jet fluid in front of the shock.  Convection of
these electrons downstream of the shock with enhanced magnetic field
results in a second synchrotron component that can explain the
anomalously hard X-ray SEDs of many knots.  Note that this component
of electrons is re-energized at the shock front in the process of
adiabatic compression.  Moreover, because of the high efficiency of
synchrotron radiation, in many cases even the energy of the neutral
beam deposited in the $\beta$-decay electrons alone could be
sufficient for the explanation of the low X-ray fluxes observed.  We
note that MHD waves excited by the beam, as in the shear boundary
layer model of \citet{so02}, could also contribute to the
reacceleration and energization of a hard electron component.

The decline of the deposition rate of UHE electrons with distance $r$
from the AGN core could be as fast as $\propto r^{-2}$ in case of a
dominant $\beta$-decay electron contribution. The decline could be
somewhat slower in the case of a significant contribution from
pair-photon cascades of UHE electrons. This would then explain the
fast decline of X-ray emission of knots along the jet.  The final hot
spot in our model would represent a termination shock of the
relativistic flow when the jet runs out of UHE neutral-beam decay
products that sustain the forward progress of the relativistic jet
into the ambient medium. Compression and additional acceleration of
the second component of electrons (cooled down to $\gamma_{min}
\gtrsim 10^{6}$ on timescales up to $1\,\rm Myr$ for hot spots at
distances $\lesssim 300 \,\rm kpc$) and the increase of the magnetic
field in the downstream region of this relativistic jet termination
shock can explain the spectrum of the hot spot of Pictor A with
$\alpha_X \geq 1$.
 
For very distant quasars with $z\gtrsim 3$, such as reported recently
\citep{sch02,siem03}, an additional powerful channel contributing to
the second electron component in large scale jet appears.  At such
redshifts, the photomeson interaction length of super-GZK neutrons
becomes comparable with and even less than the Mpc scale of the
jets. In this case, the input from the super-GZK neutron component of
the UHE neutral beam and the efficiency of the electromagnetic cascade
will increase dramatically. In this way, the increased energy density
of the CMBR can explain the detectability of X-rays even from quasars
at large redshifts (compare different ideas by
\citet{sch02a} for the EC model).

We propose two testable predictions of this scenario. The first is
variability of the X-ray flux from the knots of FR2 radio
galaxies. This possibility, suggested by observations of X-ray
variability in the knots of M87 \citep{per03}, is a consequence of the
short cooling time scale $t_{syn} \simeq 500 \sqrt{(1+z)/E_{keV}
\delta B_{30}^3}\,$yr of X-ray emitting electrons. For the comoving
magnetic fiels in the knots (and hot spots) reaching $B^\prime \gtrsim
100\,\rm \mu G$ this time can be $\lesssim 50/\sqrt{E_{keV}} \,\rm yr$
even for knots/jets with moderate Doppler factors $\delta \lesssim 3$
moving at rather large angles $\theta \lesssim 20^\circ$.  One could
then expect that variations of the injection rate of electrons at the
shock (swept up in the upstream region) on time scales $t_{var}\gtrsim
t_{syn}$ would result in variations of the X-ray flux on the same
scales.  Detection of variability could be feasible
for X-rays knots with transverse sizes $\lesssim 1\,$kpc, but which
are sufficiently thin (flat) along our line-of-sight.  
The detection of a small-amplitude variability (at
the level of up to several per cent
over reasonable observation times) variability depends on the strength
of the magnetic field downstream of the shock.  For knots detected
with {\it Chandra} at the X-ray count rates $\gtrsim 100$ per hour,
like from knot WK7.8 in PKS 0637-752 \citep{cha00}, simple estimates
show that a statistically significant flux variations could be
expected for observations separated by several years.  X-ray
variability of an FR2 knot or hot spot would be difficult to
understand in the context of the EC model with the large cooling times
of the emitting electrons and stationary target for Compton
interactions.

A second prediction is that high-energy ($\gtrsim 10^{14}$ eV)
neutrinos will be detected from core-dominated quasars with km-scale
neutrino telescopes such as IceCube.  Detection of every such neutrino
from a flat spectrum radio quasar such as 3C 279 implies that a total
energy $\sim 5\times 10^{54} \delta^{-2} \,\rm erg $ is injected in
the inner jet at sub-kpc scales \citep{ad03}. Therefore even the
detection of 1 neutrino per year would imply an UHE neutral beam power
$\sim 10^{46} \delta^{-2} \,\rm erg \, s^{-1}$.

Finally, we note that this model is developed in view of a scenario 
\citep{bd02,ce02} where radio loud AGNs are formed by merging IR
luminous galaxies, and evolve from a high luminosity FR2/quasar phase
at early cosmic times to a lower luminosity FR1/BL Lac phase at later
times, thus explaining the sequence of flat-spectrum radio quasar and
BL Lac object SEDs \citep{fos98}. The crucial feature in this scenario
is the amount of broad-line emitting gas that fuels the AGN, and the
corresponding intensities of the external radiation field in the
vicinity of the inner jet. This radiation field determines the power
of the neutral beam that produces the distinctive lobe-dominated
morphologies of FR2 radio galaxies and, in exceptional cases such as
Pictor A, a linear jet \citep{wys01}. FR1 galaxies, where the neutral
beam power is negligible, consequently display very different radio
jet morphologies.

\vskip0.5in
We thank Herman Marshall, Eric Perlman, and Andrew
Wilson for comments and discussions,  and the referee for useful comments.
AA appreciates the hospitality of the NRL High Energy Space Environment
Branch during his visit when this work has been started.
The work of CD is supported by 
the Office of Naval Research and {\it GLAST} Science Investigation
No.\ DPR-S-1563-Y.


\end{document}